\documentclass[a4paper]{jpconf}
\usepackage{graphicx}
\usepackage{amssymb}
\usepackage{amsmath}
\usepackage{cite}
\newcommand{\betaL}{\beta_{\mathrm{L}}}
\newcommand{\betaLC}{\beta_{\mathrm{L}}^{~\mathrm{c}}}

\newcommand{\gL}{g_{\mathrm{L}}}
\newcommand{\gLC}{g_{\mathrm{L}}^{~\mathrm{c}}}
\newcommand{\gLR}{g_{\mathrm{L}}^{~\mathrm{ref}}}
\newcommand{\gTC}{g_{\mathrm{T}}^{~\mathrm{c}}}
\begin{document}
%========
\title{
Thermodynamic Lattice Study
for Preconformal Dynamics in Strongly Flavored Gauge Theory}
%========
\author{Kohtaroh Miura}
\address{INFN - Laboratori Nazionali di Frascati
Via Enrico Fermi, n.40
Frascati Rome Italy}
\ead{Kohtaroh.Miura@lnf.infn.it}
%===========================
\begin{abstract}
By using the lattice Monte-Carlo simulation,
we investigate the finite temperature
chiral phase transition at color SU(3) gauge theories with
various species of fundamental fermions,
and discuss the signal of the (pre-)conformality at large
$N_f$ (num. of flavors.) via their comparisons.
With increasing $N_f$,
we confirm stronger fermion screenings
resulting from a larger fermion multiplicity.
We investigate a finite $T$ step-scaling
which is attributed to the uniqueness of
the critical temperature ($T_c$) at each $N_f$,
then the vanishing step-scaling signals
the emergence of the conformality around $N_f^* \sim 10 - 12$ .
Further, motivated by the recent
functional renormalization group analyses,
we examine the $N_f$ dependence of $T_c$,
whose vanishing behavior indicates the conformality at
$N_f^* \sim 9 - 10$.
\end{abstract}
%===========================
\section{Introduction}
%========
Conformal invariance is anticipated to
emerge in asymptotically free non-Abelian gauge theories
when the fermion species (flavor $N_f$)
exceeds a critical value $N_f={N_f}^*$
\cite{Caswell:1974gg,Banks:1981nn}.
The approach to the conformality from below
is in principle associated
with a pre-conformal (walking) behavior of the running
coupling, which has been advocated as a basis for strongly interacting
mechanisms of electroweak symmetry breaking~\cite{Sannino:2009za}.

Recent lattice studies~\cite{DelDebbio:2010zz} 
focused on the computation of ${N_f}^*$ 
and the analysis of the conformal window itself, 
either with fundamental fermions
or other representations. 
Among the many interesting results with fundamental fermions, we single
out the observation 
that the color SU(3) gauge theory with $N_f = 8$
is still in the hadronic phase
\cite{Appelquist:2009ty,Deuzeman:2008sc},
while $N_f = 12$ seems to be close to the critical number of flavors,
with some groups favoring conformality
\cite{Appelquist:2011dp,Appelquist:2009ty,
Deuzeman:2009mh,Hasenfratz:2010fi}, 
and others chiral symmetry breaking~\cite{Fodor:2011tu}.

In order to attack the walking and the conformal dynamics,
it is more informative beyond a fixed $N_f$
to investigate the vanishing or reducing chiral dynamics
with increasing varying $N_f$.
To this end,
we investigate the $N_f$ dependences of
the chiral phase transition at finite temperature ($T$)
based on our recent works~\cite{Deuzeman:2008sc,Miura:2011mc}:
The vanishing (reducing) finite T step-scaling
which attributes to the uniqueness of $T_c$ at each $N_f$ signals
the emergence of the (pre-)conformality.
Further, motivated by the recent
functional renormalization group (FRG) studies~\cite{BraunGies},
we examine the $N_f$ dependence of $T_c$,
whose vanishing (decreasing) behavior indicates
the (pre-)conformality.
This {\em thermodynamic} lattice study for
the large $N_f$ non-Abelian gauge theory
has played a crucial role to extract a notion of
more strongly interacting non-Abelian plasma~\cite{Liao:2012tw},
and it is expected to provide a new connection
between the lattice and the Gauge/Gravity duality
\cite{Gursoy:2010fj}.

%====================================
\section{Simulation setups}
%========
Simulations have been performed by utilizing
the publicly available MILC code~\cite{MILC}.
We use an improved version of the staggered action, the
Asqtad action, with a one-loop Symanzik
\cite{Symanzik}
and tadpole~\cite{LM1985} improved gauge action.
The tadpole factor $u_0$ is determined
by performing zero temperature simulations
on the $12^4$ lattice, and used as an input
for finite temperature simulations.
To generate configurations with mass degenerate
dynamical flavors,
we have used the rational hybrid Monte Carlo algorithm
(RHMC)~\cite{Clark:2006wq}.

Our observables are the chiral condensate
and the Polykov loop
%===============================
\begin{equation}
a^3\langle\bar{\psi}\psi\rangle =
\frac{N_f}{4N_s^3N_t}
\Big\langle\mathrm{Tr\bigl[M^{-1}\bigr]}\Big\rangle
\ ,\quad
L =
\frac{1}{N_cN_s^3}\sum_{\mathbf{x}}
\mathrm{Re}
\bigg\langle
\mathrm{tr}_c\prod_{t=1}^{N_t}U_{4,t\mathbf{x}}
\bigg\rangle
\ ,\label{eq:PBPLOOP}
\end{equation}
%===============================
where $N_s~(N_t)$ represents the number of lattice sites
in the spatial (temporal) direction and
$U_{4,t\mathbf{x}}$ is the temporal link variable.
The other important observable is
the ratio of
a scalar and a pseudo-scalar susceptibility~\cite{Kocic:1992is},
%===================
\begin{equation}
R_\pi \equiv \frac{\chi_\sigma}{\chi_\pi}
=
\frac{\partial \langle \bar \psi \psi\rangle/\partial m}
{\langle \bar \psi \psi\rangle/m}
=
\frac{\chi_{\mathrm{conn}} + \chi_{\mathrm{disc}}}
{\langle \bar \psi \psi\rangle/m}
\ ,\label{eq:R_pi} 
\end{equation}
%===================
where
$
a^2\chi_{\mathrm{conn}} = 
-N_f\langle \mathrm{Tr}[( MM )^{-1}] \rangle
/(4 N_s^3 N_t)
$
and
$
a^2\chi_{\mathrm{disc}}=
N_f^2
[\langle \mathrm{Tr} [M^{-1}]^2\rangle
-\langle \mathrm{Tr} M^{-1} \rangle^2]
/(16 N_s^3 N_t)
$.
Here, $R_{\pi}\sim\mathcal{O}(1)$ indicates
the scalar and pseudo-scalar degeneracy
attributed to the approximate chiral restoration,
while $R_{\pi}\ll 1$ indicates the chiral symmetry breaking
\cite{Deuzeman:2008sc,Kocic:1992is}.

%===========================
\section{Results}
%========
We evaluate the thermalized ensemble averages of
the chiral condensate (PBP) and Polyakov loop (PLOOP)
for various lattice couplings $\betaL$,
lattice sizes with the finite $T$ set up $N_s\gg N_t$,
and the number of flavors $N_f$.
All results have been obtained by using
a single value for a lattice bare fermion mass $am=0.02$.
Then we locate the lattice bare coupling $\betaLC$
associated with the chiral crossover
which is signaled by the drastic decrease (increase) of PBP (PLOOP)
as a function of $\betaL$.
In practice, the ratio of the scalar and pseudo-scalar susceptibility
$R_{\pi}$ gives a stronger signal of the chiral crossover,
owing to its renormalization invariant property.
In table \ref{Tab:bc},
we summarize the obtained critical lattice couplings
as a function of ($N_f,N_t$).
We have confirmed the (approximate) asymptotic scaling for
the normalized critical temperature
$T_c/\Lambda_{\mathrm{L}}$ varying $N_t$ at each $N_f$,
where $\Lambda_{\mathrm{L}}$ is so-called lattice Lambda.
This indicates that our $\betaLC$ have been determined
near to the continuum limit~\cite{Gupta:2000hr}.

%==========================
\begin{table*}[ht]
\caption{
Summary of the (pseudo) critical lattice couplings $\betaLC$
for the theories with $N_f=0,~4,~6,~8$, $am=0.02$
and varying $N_t=4,~6,~8,~12$.
The result has partially been extracted from our recent paper
\protect\cite{Deuzeman:2008sc,Miura:2011mc}.
All results are obtained by using the action with
the same level improvements.
}\label{Tab:bc}
\begin{center}
\begin{tabular}{c|c|c|c|c}
\hline
$N_f\backslash N_t$ &
$4$&
$6$&
$8$&
$12$\\
\hline
$0$ &
$7.35\pm 0.1$&
$7.88\pm 0.05$&
$8.20\pm 0.1$&
$-$\\
$4$ &
$5.60\pm 0.1$&
$5.89\pm 0.05$&
$6.10\pm 0.1$&
$-$\\
$6$ &
$4.65\pm 0.05$&
$5.05\pm 0.05$&
$5.2\pm 0.05$&
$5.55\pm 0.1$\\
$8$ &
$-$&
$4.1125\pm 0.0125$&
$4.20\pm 0.1$&
$4.34\pm 0.04$\\
\hline
\end{tabular}
\end{center}
\end{table*}
%========================

%========
% MY
%========
We shall now discuss the emergence of the (pre-)conformality
by using our critical lattice $\gLC = \sqrt{10/\betaLC}$ collection.
The uniqueness of a critical temperature at each $N_f$,
$
T_c^{-1} = N_t\ a(\betaLC) = N_t^{\prime}\ a({\betaLC}^{\prime})
$
with $N_t\neq N_t^{\prime}$
gives a {\em thermal}
step-scaling $\Delta\betaLC = \betaLC - {\betaLC}^{\prime}$.
A vanishing (decreasing) $\Delta\betaLC$
can be the signal of the (pre-)conformality.

\begin{figure}[ht]
\includegraphics[width=7.5cm]{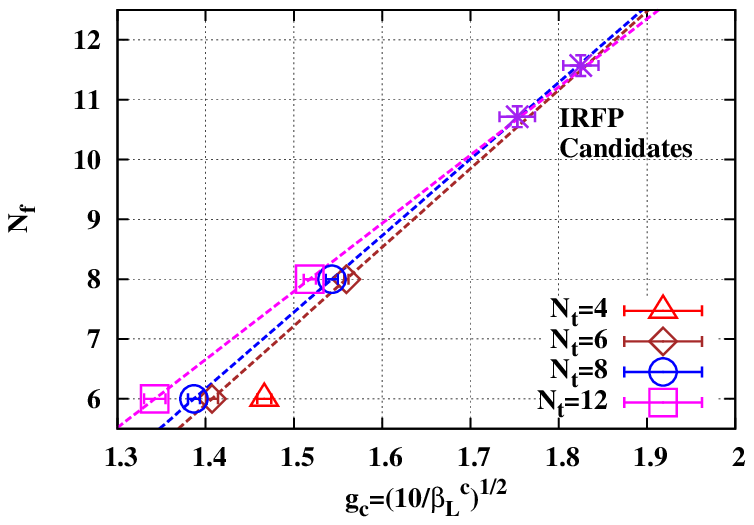}
\includegraphics[width=7.5cm]{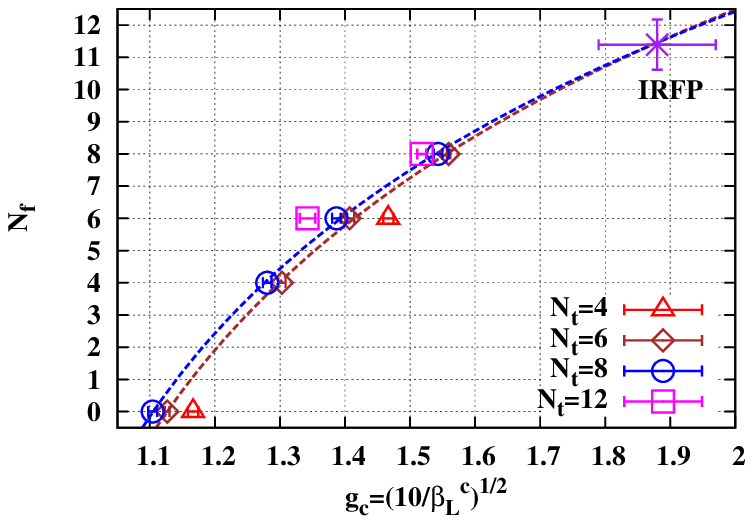}
\caption{(Pseudo) critical values of the lattice coupling
$\gLC=\sqrt{10/\betaLC}$ for theories with $N_f=0,~4,~6,~8$ 
and for several values of $N_t$
in the Miransky-Yamawaki phase diagram.
Left: We have picked up $\gLC = \sqrt{10/\betaLC}$ for
$N_f = 6$ and $8$,
and consider the ``constant $N_t$'' line
with $N_t = 6,\ 8,\ 12$. Candidates for the IRFP
have been estimated by the crossing points of
$N_t = 12$ line with $N_f = 6$ or $N_f = 8$ line.
Right:
The dashed line is a fit with the ansatz
$
N_f(\gLC) = A\cdot \log~
\bigl[B\cdot(\gLC- \gLC|_{N_f=0}) + 1\bigr]
$                        
for $N_t=6$ and $N_t=8$ results.}
\label{Fig:MY}
\end{figure}

Our thermal step-scaling is a function of
$N_f$ and two lattice temporal extensions,
$\Delta\betaLC (N_f;N_t,N_t^{\prime})$,
and tends to be smaller with increasing $N_f$.
%Here are examples obtained from Table \ref{Tab:bc}
%with the step size $s = N_t^{\prime}/N_t = 2$:
%%=========================
%\begin{align}
%&\Delta\betaLC (6;6,12) = 0.5\pm 0.12\ ,\quad
%\Delta\betaLC (8;6,12) = 0.2275\pm 0.11\\
%%
%&\Delta\betaLC (0;4,8) = 0.32\pm 0.12\ ,\
%\Delta\betaLC (4;4,8) = 0.21\pm 0.12\ ,\
%\Delta\betaLC (6;4,8) = 0.15\pm 0.075
%\ .\label{eq:Dbc}
%\end{align}
%%=========================
%The decreasing property of thermal step scalings
%indicates the pre-conformal dynamics at large $N_f$.
We here estimate the number of flavor satisfying
$\Delta\betaLC(N_f^*) = 0$
by extrapolating our $\betaLC$ collection into
the larger flavor region.
To this end, we plot the (pseudo) critical lattice coupling
$\gLC = \sqrt{10/\betaLC}$
as a function of $N_f$ in Fig.~\ref{Fig:MY},
which gives an extension of
Miransky-Yamawaki diagram \cite{Miransky:1997}
to finite $T$ cases.

Let us first pick up
the lattice critical couplings for
$N_f = 6$ and $8$,
and consider a ``constant $N_t$'' line.
As shown in the left panel of Fig.~\ref{Fig:MY},
$N_t = 6$ and $12$ lines 
get to a joint at
$(\gLC, N_f^*) = (1.825\pm 0.02,\ 11.57\pm 0.17)$,
and $N_t = 6$ and $12$ lines 
at $(\gLC, N_f^*) = (1.753\pm 0.02,\ 10.715\pm 0.17)$,
indicating the infra-red fixed point
with vanishing thermal step scalings.
Next we shall investigate
the critical lattice couplings at $N_t = 6$ and $8$
for whole range of $N_f = 0 - 8$.
They can be well fitted by assuming the functional expression
%=================================
%\begin{align}
$
N_f(\gLC) = A\cdot \log~
\Bigl[B\cdot
\bigl(\gLC
- \gLC|_{N_f=0}
\bigr) + 1
\Bigr]
$
%\ ,\label{eq:MY_fit_log}
%\end{align}
%=================================
giving $(\gLC, N_f^*) = (1.88\pm 0.09,\ 11.39\pm 0.78)$
(right panel of Fig.~\ref{Fig:MY}).
Thus, the thermal step-scaling
with the use of our lattice critical couplings
supports the emergence of conformal window
near to $12$ flavor system,
whose (pre-)conformality is now under debate
in the recent lattice studies~\cite{DelDebbio:2010zz}.

%========================
%========================

As indicated by the constant $N_t$ line in Fig.~\ref{Fig:MY},
the critical coupling is an increasing function of $N_f$
for a fixed lattice temporal extension.
This behavior is the direct consequence
of enhanced fermion screening effects
due to the larger number of fermion species.
We have observed the thermal step
(or equivalently the asymptotic) scalings
for our $\betaLC$, thereby,
the enhancement of the screening effects
gets to a physical significance.
The test of the asymptotic scaling
has been the historical homework since
the pioneering finite $T$ study at large $N_f$
by J. Kogut and his collaborators~\cite{Kogut:1985pp}
and now we have completed it.

\begin{figure}[ht]
\includegraphics[width=5.3cm]{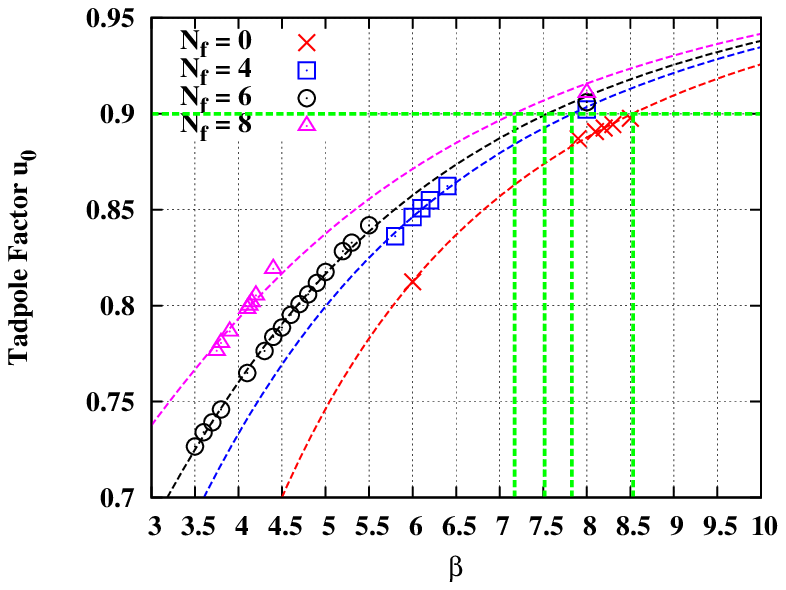}
\includegraphics[width=5.3cm]{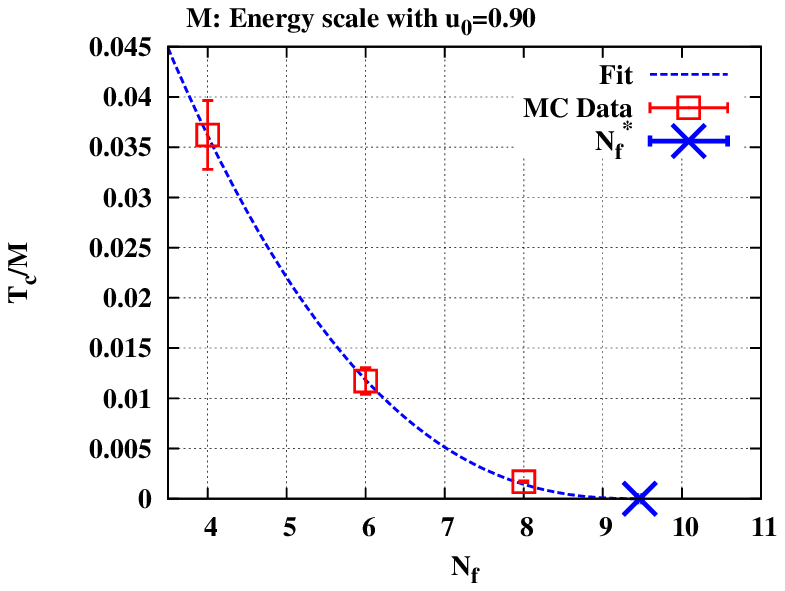}
\includegraphics[width=5.3cm]{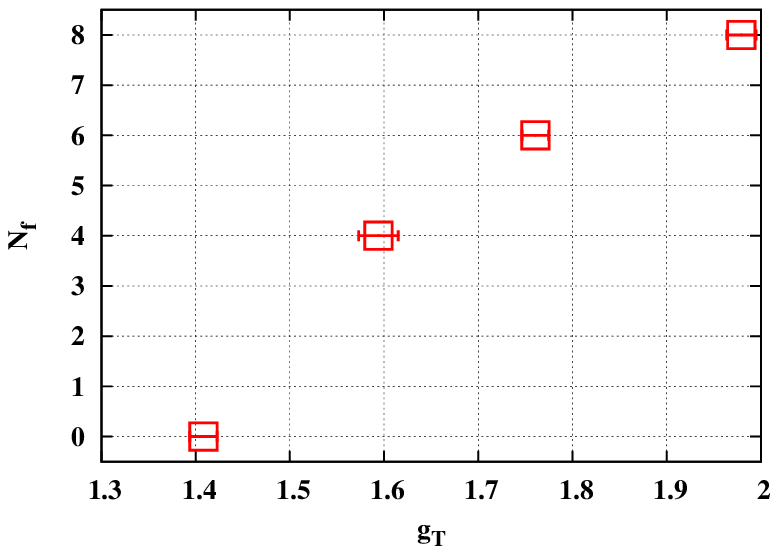}
\caption{
Left:
The $\betaL$ dependences of the tadpole factor $u_0$
at zero temperature with the use of $12^4$ lattice.
Specifying a constant $u_0$ ({\em e.g.} $u_0=0.9$ in figure),
we read off the corresponding lattice couplings $\betaL$
which are used to define the scale $M$ at each
theory with $N_f$.
Middle:
The $N_f$ dependence of $T_c/M$ where
$M$ is the UV scale with $u_0=0.9$ at each theory with $N_f$.
The dashed line represents the fit for data
by using the FRG motivated ansatz
$T_c= K|N_f^* - N_f|^{(-2b_0^2/b_1)(N_f^*)}$
Right: The thermal critical coupling
with increasing $N_f$.}
\label{Fig:TcM}
\end{figure}

%========
% Tc/M
%========
We shall now investigate the $N_f$ dependence
of the critical temperature.
In order to compare a physical quantity
such as a critical temperature
among different theories with a different $N_f$,
it is necessary to introduce
a $N_f$ independent reference scale by hand.
To set the reference scale, ideally speaking,
we would like to measure various sizes of Wilson loops
at various numbers of flavors in Monte-Carlo simulations
to obtain the running couplings $\bar{g}$
in the wide range of scales at each number of flavor.
In stead of performing such a massive simulation,
we approximately construct the renormalization flow
via the integral of the two-loop beta-function
$\beta(g)=-g^3(b_0 + b_1g^2)$
%========================
\begin{align}
\frac{T_c}{M(\gLR)}
= \frac{1}{N_t}\frac{a^{-1}(\gLC)}{M(\gLR)}
= \frac{1}{N_t}\int_{\gLR}^{\gLC}
\frac{dg}{-g^3(b_0 + b_1g^2)}
\ .\label{eq:Tc_M}
\end{align}
%========================
To specify the reference scale $M(\gLR)$,
we utilize our plaquette (tad-pole factor $u_0$) data
shown in the left panel of Fig.~\ref{Fig:TcM}.
Note that the plaquettes can be regarded as a kind of
renormalized couplings.
Let us consider a constant $u_0$ without $N_f$ dependences,
for instance $u_0 = 0.9$ in figure,
and read-off the corresponding bare lattice couplings at each $N_f$.
The obtained $\gL(N_f) = \sqrt{10/\betaL(N_f)}$ is used
as a reference coupling $\gLR$ in Eq.~(\ref{eq:Tc_M}).
This procedure imitates the scale setting
in the potential scheme renormalization,
and the use of $N_f$ independent $u_0$ is motivated by
the FRG scale setting method \cite{BraunGies}.

To be analogous to the FRG study,
we should choose $u_0$ so as to
get a UV $M(\gL)$ free from the chiral dynamics.
The middle panel of Fig.~\ref{Fig:TcM}
displays the $N_f$ dependence of
$T_c/M(\gLR)$
defined by Eq.~(\ref{eq:Tc_M}) with $u_0=0.9$.
Fitting $T_c/M(\gLR)$
with the FRG motivated ansatz
$T_c= K|N_f^* - N_f|^{(-2b_0^2/b_1)(N_f^*)}$,
we now read-off the lower edge of conformal window
$N_f^*\sim 9.47\pm 0.02$, which is somewhat smaller value
comparing to those obtained by the vanishing thermal scale settings.
In the middle panel of Fig.~\ref{Fig:TcM},
we find $T_c/M(\gLR)\ll 1$, indicating
the UV nature of the reference scale $M(\gLR)$.

To get more transparent view for the UV reference scale,
we here consider the particular reference coupling $\gLR$ -
the thermal critical coupling $\gTC$
which makes the reference scale $M(\gLR = \gTC)$
be equivalent to $T_c$ in Eq.~(\ref{eq:Tc_M}):
$T_c/M(\gTC) = 1$, giving a typical
interaction strength at $T_c$.
As shown in the right panel of Fig.~\ref{Fig:TcM},
the increasing nature of $\gTC$ indicates
a realization of
more strongly interacting non-Abelian plasma at larger $N_f$
as discussed in Ref.~\cite{Liao:2012tw}.
The criterion to set the UV reference scale
$M(\gLR)$ at every $N_f$ would be given by the condition
$\gLR\ll\gTC(N_f)$ for all $N_f$.
We find that $u_0\geq 0.84$ meets a requirement,
while the use of too large $u_0$ suffers from
the strong discritization errors.
In practice, we find that
the number of flavor giving the vanishing
$T_c/M(\gLR)$
is relatively stable within the range $0.84\leq u_0\leq 0.94$
which results in $9.85\geq N_f^*\geq 9.17$.

%========================
%========================

%===========================
\section{Summary} 
%========
We have investigated 
the (pre-)conformal dynamics 
in color SU(3) gauge theories with
multi-species of fundamental fermions
by using the lattice Monte-Carlo simulation.
In order to study the conformality
beyond the fixed number of flavor $N_f$,
we have focused on
a reducing chiral dynamics at finite $T$
as a function of increasing $N_f$.
We have observed stronger fermion screenings
resulting from a larger fermion multiplicity
at larger $N_f$.
We have investigated a finite $T$ step-scaling
which follows from the uniqueness of
critical temperature ($T_c$) at each $N_f$,
then the vanishing step-scaling signals
the conformal dynamics at $N_f^* \sim 10 - 12$.
Further, motivated by the recent
FRG based studies \cite{BraunGies},
we have examined the $N_f$ dependence of $T_c$,
by introducing a UV $N_f$ independent reference scale $M(\gLR)$.
We have used the thermal critical coupling $\gTC$ as a criterion
to insure the UV nature of $M(\gLR)$
by imposing the condition $\gLR\ll\gTC(N_f)$ for all $N_f$.
We have found that
the number of flavor giving a vanishing $T_c/M(\gLR)$
is relatively stable within the range $0.84\leq u_0\leq 0.94$,
which results in $9.85\geq N_f^*\geq 9.17$.

As a future perspective,
we should measure various sizes of Wilson loops
at various numbers of flavors,
and perform more rigorous scale settings
in the potential scheme.
It is also mandatory to investigate the chiral limit
and the thermodynamic limit at large $N_f$.
This, together with a more extended set of flavor
numbers, will allow a quantitative analysis of the critical
behavior in the vicinity of the conformal IR fixed point.

%===========================
\section*{Acknowledgments}
%=========
The author thanks Maria Paola Lombardo
and Elisabetta Pallante for
continuous discussions.
He thanks Edward Shuryak
for fruitful discussion during the xQCD workshop.
This work was in part based on the MILC Collaboration's public
lattice gauge theory code~\cite{MILC}.
The numerical calculations were carried out on the
IBM-SP6 and BG/P at CINECA,
Italian-Grid-Infrastructures in Italy,
and the Hitachi SR-16000 at YITP, Kyoto University in Japan.
%===========================
\section*{References}
%========

%===========================
\end{document}